\documentstyle{ichep}
\newcommand{\Gsim}{\mathrel{\hbox{\rlap{\lower.55ex \hbox {$\sim$}}
                   \kern-.3em \raise.4ex \hbox{$>$}}}}
\newcommand{\Lsim}{\mathrel{\hbox{\rlap{\lower.55ex \hbox {$\sim$}}
                   \kern-.3em \raise.4ex \hbox{$<$}}}}

\newcommand {\beq}    {\begin{equation}}
\newcommand {\eeq}    {\end{equation}}

\newcommand {\dSstar} {d_{{\rm S}^{\star}}}
\newcommand {\dIstar} {d_{{\rm I}^{\star}}}
\newcommand {\ES}     {E_{{\rm S}}}
\newcommand {\ESstar} {E_{{\rm S}^{\star}}}
\newcommand {\AI}     {A_{{\rm I}}}
\newcommand {\AIstar} {A_{{\rm I}^{\star}}}

\newcommand {\ncl} {non-contractible loop}

\newcommand {\sm}  {standard model}

\newcommand {\sph}   {spha\-le\-ron}

\newcommand {\ew} {electroweak}

\newcommand {\Sstar} {S$^{\star }$}
\newcommand {\Istar} {I$^{\star }$}

\newcommand {\IIbar} {I$\, \bar{{\rm I}}$}

\begin{document}
\pagestyle{empty}

\title{NEW ELECTROWEAK INSTANTON AND POSSIBLE BREAKDOWN OF UNITARITY}

\author{F. R. Klinkhamer}

\affil{Institut f\"ur Theoretische Physik, Universit\"at Karlsruhe,\\
       D--76128 Karlsruhe, Germany }

\abstract{Potential implications of a new constrained instanton solution in the
\ew ~\sm ~are discussed. Notably, there may be a
non-perturbative  unitarity violating
contribution to the total cross-section at high collision energies.}

\twocolumn[\maketitle]

\section{Introduction}

In the last year we have given for the \ew ~\sm ~:
\vspace{1\baselineskip} \newline
1. an explicit construction of a new static, but unstable, classical
solution, the \sph ~\Sstar ~\cite{K93b} ;
\vspace{0\baselineskip} \newline
2. an existence ``proof'' (and construction method) for the related
constrained instanton \Istar ~\cite{K93a} .
\vspace{1\baselineskip} \newline
Roughly speaking, \Sstar ~is a constant time slice 
through \Istar, just like the well-known \sph ~S \cite{KM84} resembles
a constant time slice of the BPST instanton I \cite{BPST75}.
\vspace{0\baselineskip} \newline
The outline of this talk is as follows.
First we recall a few pertinent facts about these two new classical solutions.
Then we make some remarks on the potential physics implications,
focussing on the role of the new instanton \Istar.
Specifically, we mention the asymptotics of perturbation theory
and the apparent violation of unitarity at high energies.

\section{Classical solutions}

\noindent The \sph ~\Sstar ~has the following characteristics \cite{K93b} :
\vspace{1\baselineskip} \newline
1.  axial symmetry of the fields ; \newline
2.  vanishing Higgs field at two points on the symmetry axis, separated by
    a distance $\dSstar$ ; \newline
3.  chiral fermion zeromodes (related to the global $SU(2)$ anomaly),
    localized at either point of vanishing Higgs field,
    depending on the chirality ; \newline
4.  energy $\ESstar \sim 2\,\ES \sim 20\,{\rm TeV}$ and
      $\dSstar \sim 4\,M_{W}^{-1}$ .
\vspace{1\baselineskip} \newline
The instanton \Istar ~has not yet been constructed in all detail, but at least
the following properties are clear \cite{K93a}, starting with a
technical preliminary :
\vspace{1\baselineskip} \newline
0. constraint needed to fix the scale ($\rho$) of the solution ; \newline
1. axial symmetry of the fields ($U(1)$-equivariance) ; \newline
2. Higgs zeros seperated by a distance $\dIstar$ ; \newline
3. localized chiral fermion zeromodes ; \newline
4. action $\AIstar \sim 2 \, \AI
      \sim (16 \, \pi^{2} / g^{2})
      \left( 1 + {\rm O}(\rho^{2} M_{W}^{2}) \right)$ and
   distance parameter
   $\dIstar \sim M_{W}^{-1}\left( 2 + {\rm O}(\rho\, M_{W})\right)$ ; \newline
5. resemblance to a very loose di-atomic molecule for scales
   $\rho << M_{W}^{-1}$ .

\section{Perturbation theory}

The instanton ~\Istar ~sits at the top of a \ncl ~of 4-dimensional
euclidean configurations and is assumed to have only one negative mode.
In a way \Istar ~is like the \sph ~of a 5-dimensional theory.
(Note that \Istar ~is not a ``bounce'' solution \cite{C77}, because of the
absence of a ``turning point'' with vanishing field derivatives.)
\vspace{1\baselineskip} \newline
Following Lipatov \cite{L77} we see that \Istar ~determines the
a\-symp\-to\-tics of \ew ~perturbation theory
\beq
c_{k}\, g^{2k} \propto  \frac{k !}{(\AIstar)^{k}}
             \sim     \frac{k !}{(16 \, \pi^{2})^{k}}\,g^{2k}\; ,
\label{eq:asymptotics}
\eeq
where $c_{k}$ are the coefficients calculated for an arbitrary Green's
function.  The same behaviour has actually
been verified for the groundstate energy
in a quantum mechanical model \cite{RS94}.
\vspace{1\baselineskip} \newline
The result (\ref{eq:asymptotics}) on the asymptotics
can be rephrased \cite{tH79} by saying that
the solution \Istar ~gives a singularity at $ z \sim 16 \, \pi^{2} $
in the Borel plane (variable $z$ corresponding to $g^{2}$).

\section{Unitarity}

A straightforward calculation  \cite{K93a,K92}
of the two-fermion forward elastic scattering (FES) amplitude
gives from the euclidean
path integral (the dominant contribution being close to $\rho = 0$)
\beq
F(s,0)^{\rm non-pert} \propto
         \exp\left[\; \sqrt{s}\,\dIstar(0) - \AIstar(0)\; \right]\; ,
\label{eq:fes}
\eeq
with $F(s,t)$ the scattering amplitude as a function of the
Mandelstam variables.      
This behaviour follows from inserting the \Istar ~fields into the euclidean
path integral for the 4-point Green's function and integrating over the
collective coordinates.  The first term in the exponent (\ref{eq:fes})
comes from the Fourier transform of the
fermion zeromodes, which are asymmetric with a distance parameter
$\dIstar$, and the analytic continuation from euclidean to
minkowskian space-time. The second term is simply the instanton action.
\vspace{1\baselineskip} \newline
Remark that our calculation is similar to an earlier one with
an {\em approximate\/} solution \IIbar ~\cite{P90,KR91}.
We use, instead, the only known {\em exact\/} solution (\Istar)
relevant to the problem.
\vspace{1\baselineskip} \newline
The non-perturbative contribution (\ref{eq:fes}) is generic
to all FES amplitudes and violates unitarity at
\beq
\left( \sqrt{s} \right)_{\rm threshold}
      \sim \frac{\AIstar(0)}{\dIstar(0)}
      =    \left( \frac{\AIstar(0)}{16 \, \pi^{2}/g^{2}}    \right)
           \left( \frac{2 \, M_{W}^{-1}}{\dIstar(0)} \right)
           \tilde{\ESstar} \; ,
\label{eq:threshold}
\eeq
with the definition
\[ \tilde{\ESstar} \equiv 2\, \pi\, M_{W} / \alpha_{w} \sim \ESstar \; .\]
The point is that by the optical theorem (unitarity)
the imaginary part of the FES amplitude should
be related to the total cross-section,
with the Froissart bound (unitarity and
analyticity) $\sigma^{\rm total} < {\rm O}(\log^{2} s) $, and
this bound is rapidly violated by
the exponential increase with center of mass energy $\sqrt{s}$ as given
by (\ref{eq:fes}).
More directly, the exponential behaviour (\ref{eq:fes})
violates the polynomial boundedness condition
$ | F(s,0) | < s^{N} \; $,
with $N$ a finite power, see for example \cite{E67}.

\vspace{1\baselineskip}
\noindent Clearly this is a serious problem for \ew ~field theory which
{\em must\/} be solved. We see three possible solutions :
\vspace{1\baselineskip}
\newline
1. unitarity restoration together with the Feynman per\-tur\-ba\-ti\-on series
;
\newline
2. inapplicability of the conventional euclidean path integral formalism, cf.
   \cite{V94} ; \newline
3. modification of the \sm ~.
\vspace{1\baselineskip}
\newline
Elsewhere we hope to elaborate on the first, most conservative, alternative.
Here we only remark that this possible solution
may provide us with additional constraints on the parameters of the theory.
\vspace{1\baselineskip} \newline
If, however, there is a significant B+L violating part to $\sigma^{\rm total}$
from (\ref{eq:fes}), then solutions 2 and/or 3  may be forced upon us.

\section{Conclusions}

New classical solutions of the \ew ~field
equations have been discovered recently,
the \sph ~\Sstar ~and the instanton \Istar. In this talk we have argued that
these classical solutions play a role in some of the most
fundamental questions of \ew ~field theory,
viz. the meaning of perturbation theory and unitarity.

\section*{Acknowledgements}

The author would like to acknowledge the continued hospitality of
CHEAF and NIKHEF-H.
He also thanks M. Veltman for valuable discussions.

\end{document}